\documentclass[aps,pra,twocolumn,superscriptaddress]{revtex4-1}

\usepackage{blindtext}
\usepackage{graphicx}
\usepackage{epstopdf}
\usepackage{amssymb}
\usepackage{amsmath}
\usepackage{mathtools}
\usepackage{bm}
\usepackage{xcolor}

\begin{document}

\title{Observation of wall-vortex composite defects in a spinor Bose-Einstein condensate}

\author{Seji Kang}
\affiliation{Department of Physics and Astronomy, and Institute of Applied Physics, Seoul National University, Seoul 08826, Korea}
\affiliation{Center for Correlated Electron Systems, Institute for Basic Science (IBS), Seoul 08826, Korea}

\author{Sang Won Seo}
\affiliation{Department of Physics and Astronomy, and Institute of Applied Physics, Seoul National University, Seoul 08826, Korea}

\author{Hiromitsu Takeuchi}
\affiliation{Department of Physics and Nambu Yoichiro Institute of Theoretical and Experimental Physics (NITEP), Osaka City University, Osaka 558-8585, Japan}

\author{Y. Shin}
\email{yishin@snu.ac.kr}
\affiliation{Department of Physics and Astronomy, and Institute of Applied Physics, Seoul National University, Seoul 08826, Korea}
\affiliation{Center for Correlated Electron Systems, Institute for Basic Science (IBS), Seoul 08826, Korea}

\begin{abstract}
We report the observation of spin domain walls bounded by half-quantum vortices (HQVs) in a spin-1 Bose-Einstein condensate with antiferromagnetic interactions. A spinor condensate is initially prepared in the easy-plane polar phase, and then, suddenly quenched into the easy-axis polar phase. Domain walls are created via the spontaneous $\mathbb{Z}_2$ symmetry breaking in the phase transition and the walls dynamically split into composite defects due to snake instability. The end points of the defects are identified as HQVs for the polar order parameter and the mass supercurrent in their proximity is demonstrated using Bragg scattering. In a strong quench regime, we observe that singly charged quantum vortices are formed  with the relaxation of free wall-vortex composite defects. Our results demonstrate a nucleation mechanism for composite defects via phase transition dynamics.
\end{abstract}

\maketitle

Topological defects in a continuous ordered system are a splendid manifestation of symmetry breaking, with their fundamental types, such as walls, strings, and monopoles, inevitably determined by the topology of the order parameter space. However, if there is a hierarchy of energy (length) scales with different symmetries, composite defects, such as domain walls bounded by strings and strings terminated by monopoles may exist in the system~\cite{Kibble_book}. In cosmology,  it has been noted that such composite defects can be nucleated through successive phase transitions with different symmetry breaking in grand unification theories; furthermore, composite defect formation has been proposed as a possible mechanism for galaxy formation~\cite{Kibble_book,Vilenkin_phyrep85} and baryogenesis~\cite{Menahem_nuclphys92} in the early Universe.  

Spinful superfluid systems with multiple symmetry breaking provide an experimental platform for studying the physics of composite defects and, thus, to examine the cosmological scenario. In superfluid $^3$He-B, it has been observed that a spin-mass vortex, on which a planar soliton terminates, can survive after phase transitions by being pinned on the vortex lattice~\cite{Kondo_prl92,Eltsov_prl00} or nafen~\cite{Makinen_arxiv18}. Composite defects have also been theoretically studied in the atomic Bose-Einstein condensate (BEC) system. Vortex confinement with a domain wall was predicted to occur in a two-component BEC under coherent intercomponent coupling~\cite{Son_pra02}. In particular, for a spin-1 Bose gas with antiferromagnetic interactions, half-quantum vortices (HQVs) joined by a spin domain wall were anticipated to be responsible for the emergence of an exotic 2D superfluid phase with spin-singlet pair correlations~\cite{Mukerjee_prl06,James_prl11}.

In this Letter, we report the experimental observation of wall-vortex composite defects in a quasi-2D antiferromagnetic spin-1 BEC. The composite defects are nucleated via a two-step instability mechanism in quantum quench dynamics from the easy-plane polar (EPP) phase into the easy-axis polar (EAP) phase. In the first step, spontaneous $\mathbb{Z}_2$ symmetry breaking causes domain wall formation, the core of which is occupied by the EPP phase. In the second step, the snake instability splits the domain walls into segments, with each segment forming a composite defect, which is a domain wall terminating on a HQV~\cite{Seo_prl15}. The mass supercurrent in proximity to the wall end point is demonstrated using Bragg scattering~\cite{Seo_scirep17}. We also observe that singly charged quantum vortices (QVs) can be formed by the relaxation of free composite defects. Our results directly demonstrate the existence of wall-vortex composite defects and their nucleation mechanism via phase transition dynamics in a spinful superfluid system.

The experiment is performed with a BEC of $^{23}$Na atoms in the $F=1$ hyperfine state having an antiferromagnetic spin interaction coefficient $c_2>0$~\cite{Stenger98}. The ground state of a spin-1 antiferromagnetic BEC is a polar state with $\langle \mathbf{F}\rangle=0$~\cite{Ho98,Ohmi98}, where $\mathbf{F}=(F_x, F_y, F_z)$ is the spin operator of the particle. The order parameter of the BEC is parametrized with the superfluid phase $\phi$ and a real unit vector $\hat{\bm{d}}=(d_x, d_y, d_z)$ for the spin director, and is expressed as 
\begin{equation}\label{eq:order}
\bm{\psi}=\begin{bmatrix}
\psi_{+1}\\ \psi_0 \\ \psi_{-1}
\end{bmatrix}= \sqrt{n}e^{i\phi}\begin{bmatrix}
-\frac{d_x-id_y}{\sqrt{2}} \\
d_z  \\
\frac{(d_x+id_y)}{\sqrt{2}}
\end{bmatrix},
\end{equation}
where $\psi_{m_z=0,\pm1}$ is the condensate wave function of the $|m_z\rangle$ Zeeman component and $n$ is the particle density. In the presence of an external magnetic field, e.g., along the $z$ axis, uniaxial spin anisotropy is imposed by the quadratic Zeeman energy $E_z=q(1-d_z^2)$ and the ground state of the system is the EAP state with $\hat{\bm{d}}=\pm \hat{\bm{z}}$  for $q>0$ and the EPP state with $\hat{\bm{d}}\perp\hat{\bm{z}}$ for $q<0$.

\begin{figure}
	\centering
	\includegraphics[width=0.95\linewidth,keepaspectratio]{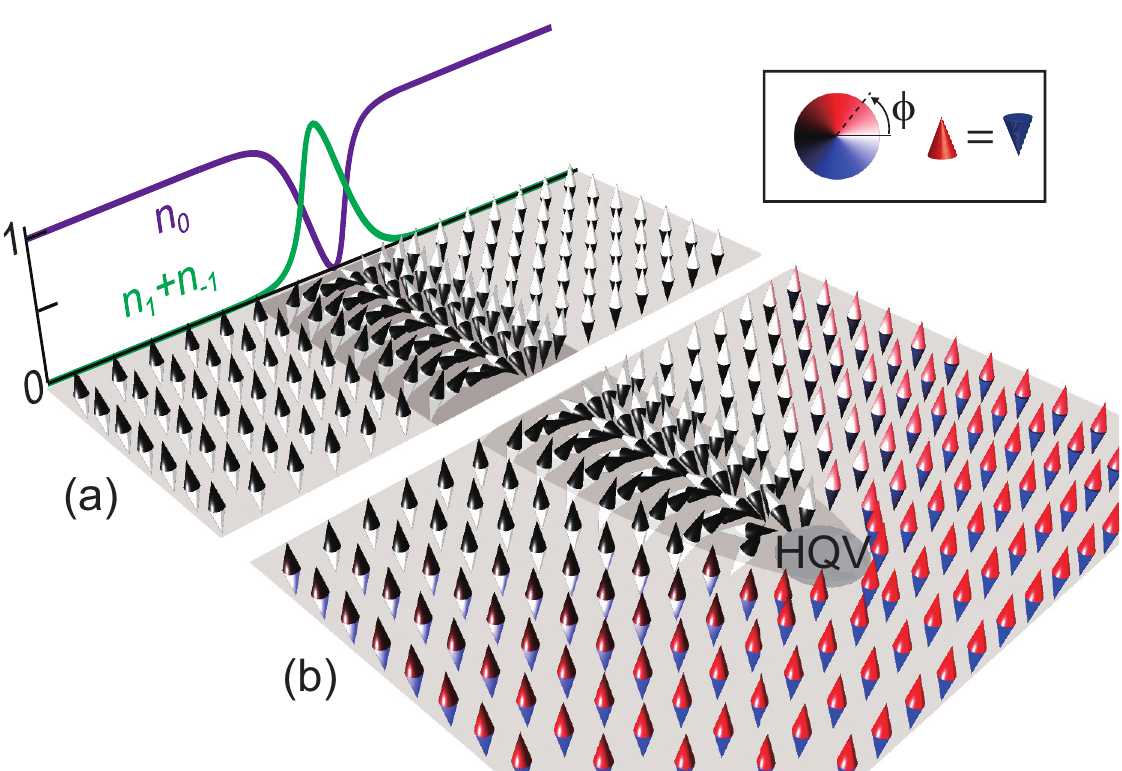}
	\caption{Schematic illustration of the wall-vortex composite defect in the EAP phase of an antiferromagnetic spinor BEC. The order parameter of the polar phase has a discrete symmetry under the operation of  $(\phi, \hat{\bm{d}})\rightarrow(\phi+\pi, -\hat{\bm{d}})$. The double-head arrow denotes the spin director $\hat{\bm{d}}$ and the color of each arrow head indicates the superfluid phase $\phi$. (a) Domain wall at the interface of two domains with opposite spin directions. $\hat{\bm{d}}$ flips to the opposite direction across the wall. The density distributions $n_{0,\pm1}$ of the three $m_z=0,\pm1$ spin components are displayed. (b) Domain wall bounded by a HQV. As the two domains are continuously connected to each other with changing $\phi$ by $\pi$, the domain wall spatially terminates and a HQV is formed at the wall end point.}
\label{fig:Phase}
\end{figure}

As a means of creating wall-vortex composite defects, we employ the quantum quench dynamics from the EPP phase to the EAP phase via a sudden change of spin anisotropy, which can be implemented by dynamically controlling the $q$ value~\cite{Gerbier_pra06,Zhao_pra14}. Because of the positional difference between the two phases in the order parameter space, the quench dynamics involves spontaneous $\mathbb{Z}_2$ symmetry breaking, as $\hat{\bm{d}}$$\perp$$\hat{\bm{z}}\rightarrow\hat{\bm{d}}=$$\pm\hat{\bm{z}}$. For $q$$>$0, the initial EPP state is dynamically unstable  so that spin fluctuations will be exponentially amplified via the spin exchange process of $|$$+1\rangle |$$-$$1\rangle \rightarrow |0\rangle|0\rangle$~\cite{Kawaguchi_phyrep12}. The microscopic origin of the $\mathbb{Z}_2$ symmetry breaking arises from the two equivalent choices for the phase of the $|0\rangle$ component. A rapid quench can give rise to a complex network of domain walls in a uniform system according to the Kibble-Zurek mechanism~\cite{Kibble80,Zurek85}. 

The spatial structure of a domain wall is described in Fig.~\ref{fig:Phase}(a), which is formed at the interface between two domains with opposite spin directions. Here $\hat{\bm{d}}$ is denoted by a double-head arrow and the superfluid phase is indicated by the color of each arrow head, reflecting the discrete symmetry of the order parameter under the operation of $(\phi, \hat{\bm{d}})\rightarrow(\phi+\pi, -\hat{\bm{d}})$~\cite{Zhou_prl01}.  In the wall region, $\hat{\bm{d}}$ continuously flips to the opposite direction and the $|$$\pm 1$$\rangle$ components are present, sandwiched by the $|0\rangle$ component. The wall thickness is determined by the competition between the quadratic Zeeman energy and the gradient energy associated with the vector field $\hat{\bm{d}}(\bm{r})$, giving a characteristic length scale of $\xi_q=\hbar/\sqrt{2mq}$ for $q\ll \mu$ with the particle mass $m$ and the chemical potential $\mu$~\cite{supple}. An interesting observation is that the two domains separated by the wall comprise only the $|0\rangle$ component, which means that they can be continuously connected to each other by varying $\phi$ without flipping $\hat{\bm{d}}$, thus, allowing spatial termination of the domain wall as shown in Fig.~\ref{fig:Phase}(b). In this case, the wall end point exhibits a superfluid phase winding of $\pi$, forming a HQV~\cite{Seo_prl15}. This is the wall-vortex composite defect expected in the EAP phase. The spatial structure of the composite defect is analogous to that of the spin-mass vortex, also referred to as a $\theta$ soliton, in superfluid $^3$He~\cite{Kondo_prl92,Makinen_arxiv18,supple}.

\begin{figure}
	\centering
	\includegraphics[width=1.0 \linewidth,keepaspectratio]{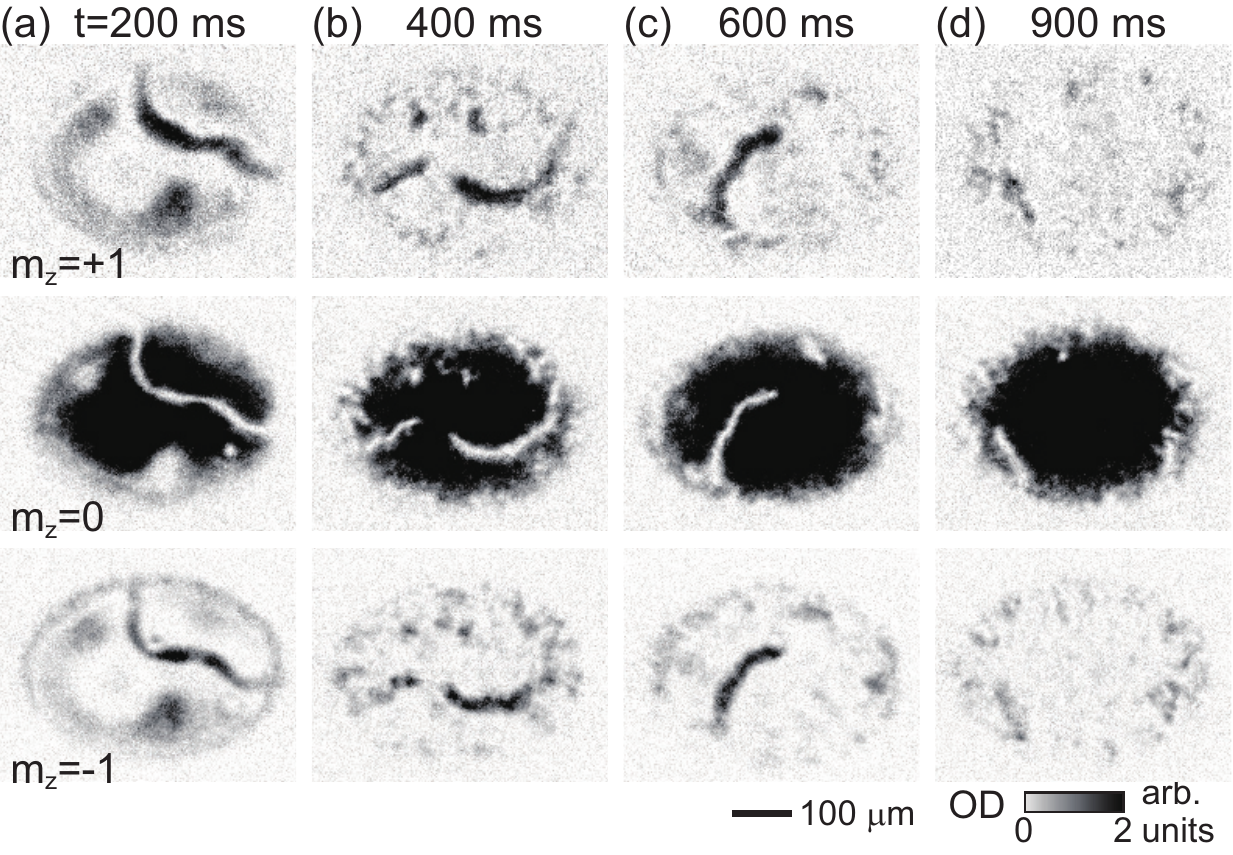}
	\caption{Creation of wall-vortex composite defects in a BEC via quantum quench into the EAP phase. Time-of-flight absorption images of the three spin components of the BEC at various hold times $t$ after the quench for $q_f/h=1.0~$Hz. Density-depleted lines in the $m_z=0$ component represent the position of domain walls, which are filled by the $m_z=\pm 1$ components.}\label{fig:Density}
\end{figure}

We prepare a condensate containing $N_c\approx 8.0\times 10^6 $ atoms  in the $|F$$=$$1,m_F$$=$$0\rangle$ hyperfine spin state in an optical dipole trap with trapping frequencies of $(\omega_x,\omega_y,\omega_z)=2\pi\times(3.8,5.5,402)~$Hz. The Thomas-Fermi radii for the trapped condensate are $(R_x, R_y, R_z)\approx(230,160,2.2)~\mu$m. The external magnetic field is $B_z=33$~mG, giving $q/h= 0.3$~Hz, and the field gradient is controlled to be less than 0.1~mG/cm~\cite{Kang_pra17}. The EPP-to-EAP quench dynamics is initiated by rotating $\hat{\bm{d}}$ from $\hat{\bm{z}}$ to the $xy$ plane by applying a short rf pulse and then suddenly changing the $q$ value to a target value $q_f>0$ using a microwave dressing technique~\cite{Seo_prl15}. 

The postquench evolution of the BEC is examined by measuring the spatial density distributions of the three spin components at a variable hold time $t$ with taking an absorption image after Stern-Gerlach (SG) spin separation for 24~ms time-of-flight~\cite{Kang_pra17}. The spin healing length is $\xi_s=\hbar/\sqrt{2mc_2 n_0}\approx 4.0~\mu$m for the peak atom density $n_0$, and our highly oblate sample with $R_z<\xi_s$ constitutes a quasi-2D system for spin dynamics. In our experiment, $q_f$, which represents the initial excitation energy per particle with respect to the ground state, is much smaller than $\mu\approx h\times 880~$Hz, so incurrence of density perturbations is energetically improbable. Note that $q_f\ll \mu$ sets a clear hierarchy of energy scales in the system~\cite{supple}.

Figure~\ref{fig:Density} shows several images of the quenched condensate after various hold times for $q_f/h=1.0$~Hz. In the early stage of the quench dynamics, a few line defects are clearly observed to appear across the condensate [Fig.~\ref{fig:Density}(a)]. The $|0\rangle$ component shows density-depleted trenches and the trench regions are filled by both of the $|$$\pm$$1\rangle$ components, consistent with the spin distribution for the domain wall described in Fig.~1(a). The high visibility of the trench in the $|0\rangle$ component after such a long time-of-flight reflects the nature of a topological soliton. At later $t>0.2$~s, the line defects are typically observed to end in the middle of the condensate [Figs.~\ref{fig:Density}(b) and \ref{fig:Density}(c)]. The profile of the total condensate density was confirmed to remain unperturbed by imaging without SG spin separation. Such a smooth wall termination is the key characteristic of the wall-vortex composite defects. As $t$ increases, the end point seems to recede toward the boundary of the condensate with the decrease of the domain wall length [Fig.~2(d)], which we attribute to the wall tension due to the quadratic Zeeman energy. It is noteworthy that ring-shaped domain walls were sporadically observed~\cite{supple}, which is reminiscent of a 2D skyrmion that is also a topologically allowed defect for the polar phase~\cite{Choi_prl12}.

To corroborate the existence of HQVs at the wall end points, we measure the mass superflow distribution using a spatially resolved Bragg scattering method~\cite{Seo_scirep17}. Before applying a pulse of magnetic field gradient for the SG spin separation, we irradiate two pairs of counterpropagating Bragg laser beams onto the sample in the $xy$ plane for $0.8$~ms. The frequencies of the laser beams are set to be resonant to atoms with a velocity of 0.3~mm/s $\approx 0.4 \frac{\hbar}{m \xi_s}$ so that the atoms that have such high velocities near the HQV cores may be scattered out of the condensate. The HQV core size is characterized by $\xi_s$~\cite{Seo_prl15,Anglin03,Lovegrove12}. Then, the mass circulation around the HQVs can be identified by examining the spatial distribution of the scattered atoms with respect to the wall end points~\cite{supple,footnote:Condition}.

\begin{figure}
	\centering
	\includegraphics[width=0.85\linewidth,keepaspectratio]{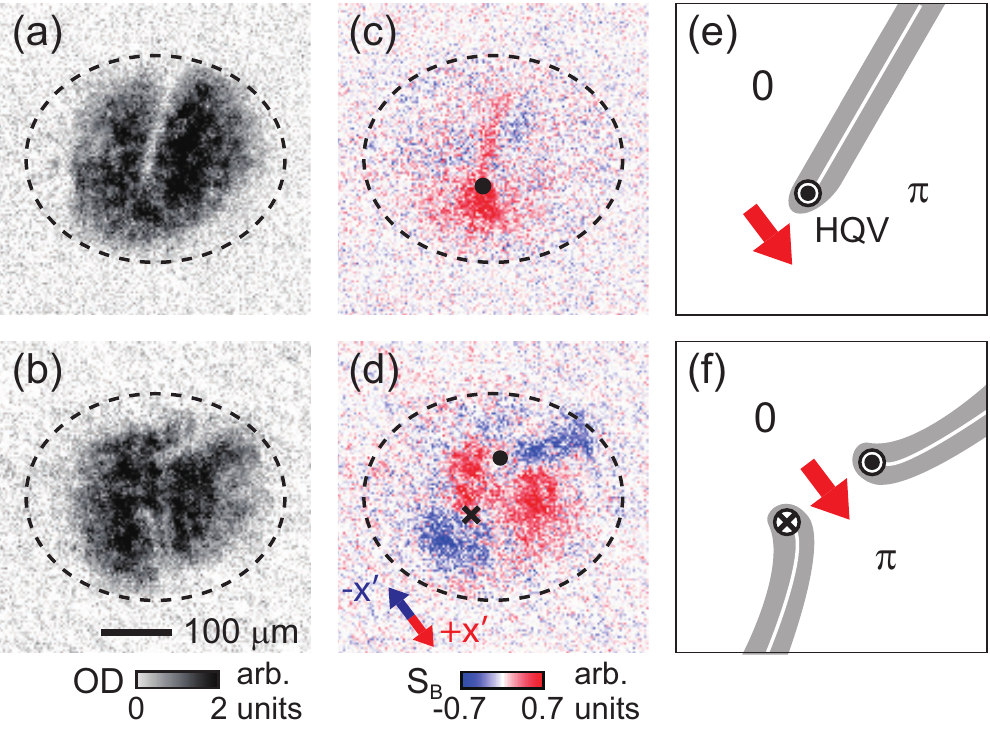}
	\caption{Detection of superflow near the defect end points by using Bragg scattering~\cite{Seo_scirep17}. (a,b) Absorption images of the $m_z=0$ component at $t=600$~ms after quench for $q_f/h=2.6~$Hz and (c,d) the corresponding images of the Bragg signal $S_B$~\cite{supple}, where the color indicates the direction of the superflow along the Bragg scattering axis ($x'$). The $\bullet$ ($\times$) mark denotes the counterclockwise (clockwise) circulation directions of the HQVs at the wall end points. The dashed lines indicate the boundary of the whole condensate. (e,f) Schematic description of the composite defect configurations corresponding to (a) and (b), respectively. The domain wall region is denoted by a grey area with a white center line and the superflow direction is indicated by a red arrow.}\label{fig:Bragg}
\end{figure}

Two examples of data of the Bragg scattering measurement are provided in Fig.~\ref{fig:Bragg}. In the case of Fig.~\ref{fig:Bragg}(a), a single domain wall terminates in the center region and a strong scattering signal is detected at the front side of its end point, consistent with the mass flow expected from a HQV with counterclockwise circulation at the end point~[Fig.~\ref{fig:Bragg}(e)]. Figure~\ref{fig:Bragg}(b) presents another case in which two end points are close to each other. The Bragg signal shows that a superflow passes through the gap between the two walls, indicating that the two HQVs at their end points have opposite circulations~[Fig.~\ref{fig:Bragg}(f)]. The spatial configuration of the two walls, together with the superflow pattern, conjures up the possibility that they might be formed by breaking a single domain wall that initially traverses the condensate [Fig.~2(a)].

The domain wall can be viewed as a three-component soliton~\cite{Nistazaki_pra08,Xiong_pra10, Bersano_prl18}, where a dark soliton of the $|0\rangle$ component with a $\pi$ phase step coexists with the bright solitons of the $|$$\pm$$ 1\rangle$ components. A dark soliton in a scalar BEC is dynamically unstable due to snake instability, which causes a local Josephson current by breaking the soliton~\cite{Feder_pra00,Anderson_prl01,Huang_pra03}. If a similar mechanism is active for the domain wall, the wall can break into many free composite defects, i.e., domain walls bounded by two HQVs at both ends. In the experiment with $q_f/h=1.0$~Hz, however, we rarely observed free composite defects detached from the condensate boundary, which means that the domain wall does not suffer much from the snake instability. It is the $|$$\pm1$$\rangle$ components at the wall core that suppress the snake instability by providing an effective pinning potential, which, thus, stabilizes the domain wall.  In other words, in the quench dynamics for large $q_f$, the snake instability can be enhanced with the domain wall becoming thinner, thus leading to proliferation of free composite defects.

\begin{figure}
	\centering
	\includegraphics[width=1.0\linewidth,keepaspectratio]{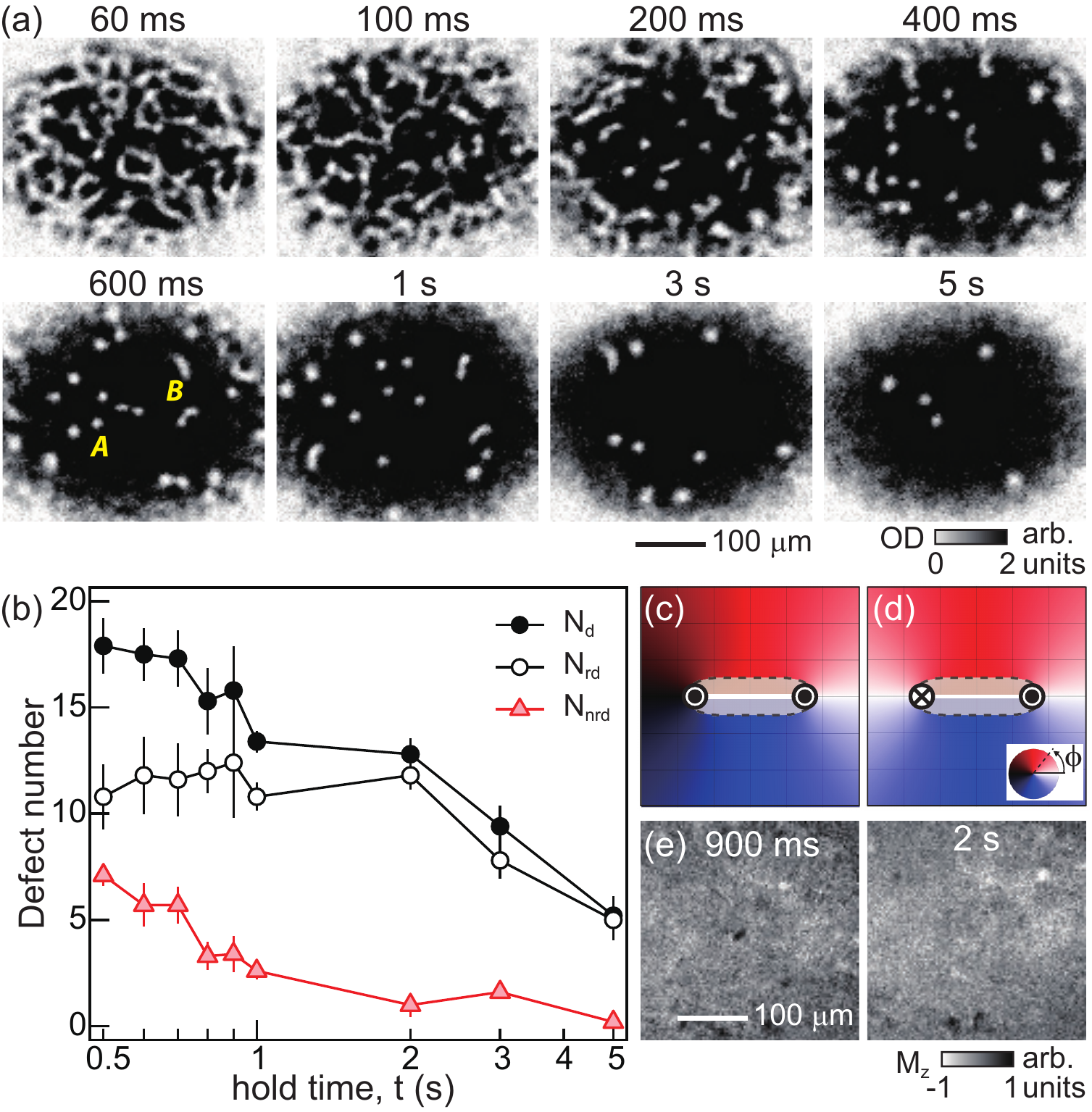}
	\caption{Temporal evolution of wall-vortex composite defects in the quenched BEC for $q_\text{f}/h=10.6~$Hz. (a) Absorption images of the $m_z=0$ component taken at various hold times $t$ after the quench. A complex network of domain walls is nucleated and the walls dynamically split into many composite defects. (b) Free defect number $N_d$ as a function of $t$. $N_\textrm{rd}$ denotes the number of round defects, whose shape aspect ratio is $<1.2$, and $N_\textrm{nrd}=N_d-N_\textrm{rd}$.  Each data point was obtained from five measurements of the same experiment and its error bar denotes their standard deviation. (c,d) Schematic description of free composite defects with different HQV configurations. The superfluid phase winding around the free defect is (c) $2\pi$ or (d) $0$. (e) \textit{In situ} images of the axial magnetization $M_z$ in the quenched BEC at $t=900$~ms (left) and 2~s (right), showing long-lived magnetized defects.}\label{fig:DomainWall}
\end{figure}

We perform the same quench experiment with a higher $q_f/h=10.6$~Hz. The Bogoliubov analysis of the dynamic instability of the initial EPP state shows that the characteristic length and time scales for the quench dynamics are proportional to $\sqrt{q_f}$ and $1/\sqrt{q_f}$, respectively, for $q_f\ll c_2n_0$~\cite{Kawaguchi_phyrep12}, and it is expected that domain walls will be nucleated in a denser pattern within a faster time scale. Indeed, we observe that a characteristically dense network of thinner domain walls develops within 60~ms and, also, that the domain walls subsequently break into many composite defects [Fig.~\ref{fig:DomainWall}(a)], demonstrating the enhancement of the snake instability.

After the fast wall splitting process, short free defects are clearly identified in the center region of the condensate at $t>0.2$~s. The free defects have a spatial size $\lesssim 5 \xi_s$ with various shapes. Some of the defects appear very round [Fig.~4(a)A], while others show dumbbell shapes, implying splitting [Fig.~4(a)B]. In the subsequent relaxation evolution, the defect number $N_d$ decays with a $1/e$ lifetime of $\approx 5$~s, where it is observed that most long-lived free defects exhibit round shapes [Fig.~4(b)]. At $t>0.8$~s, the fractional population of the round defects increases to over 80\%. Here we count a defect as a round one if its aspect ratio ($\geq 1$) is smaller than $1.2$. From \textit{in situ} magnetization imaging~\cite{Seo_prl15}, we find that the long-lived defects can have nonzero axial magnetization, i.e., contain unequal $|$$\pm $$1\rangle$ spin populations at their cores~[Fig.~\ref{fig:DomainWall}(e)], which indicates that the spin current dynamics is intricately involved in the defect formation process.

The free wall-vortex composite defects are classified into two types according to the net circulation around them. One type has a circulation of $h/m$ with two HQVs having the same circulation, which is topologically identical to a singly charged QV in a coarse-grained view [Fig.~\ref{fig:DomainWall}(c)], and the other involves with two HQVs having opposite circulations, which might be described as a magnetic bubble having linear momentum [Fig.~\ref{fig:DomainWall}(d)]. We refer to them as vortex-vortex (VV) and vortex-antivortex (VAV) types, respectively. Immediately after domain wall splitting, the system can contain statistically equal numbers of the defects for the two types. However, when the free defects become short, comparable to $\xi_s$ as observed, the defect dynamics of each type will be different because of different HQV interactions~\cite{Seo_prl16}. We may expect that small VV-type defects will survive longer with the topological character of singly charged QVs, whereas those of the VAV type will decay faster due to their linear motion in the trapped sample and possible vortex pair annihilation~\cite{Kwon14}. In the experiment, we identify the long-lived round defects as the VV type by confirming the superflow circulation around them with the Bragg scattering measurement~\cite{supple,Lovegrove14}.

Finally, we remark on the peculiarity of defect formation in the EPP-to-EAP quantum phase transition. Since the U(1) symmetry is simply broken in the final EAP ground state in a similar manner to the case of scalar BECs, one may expect a conventional Kibble-Zurek scenario for vortex nucleation in our system. However, we observe a two-step defect formation process: first, domain wall creation via the $\mathbb{Z}_2$ symmetry breaking, and then, production of composite defects by a splitting of the domain walls. The two-step scenario is also confirmed in our numerical simulation for a uniform system, based on the Gross-Pitaevskii equation for spin-1 BECs~\cite{supple,Takeuchi}. Our observations show that defect formation in phase transition dynamics critically depends on the symmetry breaking sequence of the system~\cite{Vilenkin_phyrep85}.

In conclusion, we observed the creation of wall-vortex composite defects in the EPP-to-EAP quantum quench dynamics of an antiferromagnetic BEC and demonstrated the unconventional mechanism of defect formation in the phase transition dynamics. Our findings provide a different framework for the nucleation of composite defects via the Kibble-Zurek mechanism~\cite{Kibble_book, Eltsov_prl00}. Additionally, the observation of the free composite defects encourages the efforts to search for the exotic superfluid phase in 2D antiferromagnetic spinor gases~\cite{James_prl11, Chandrasekharan_prl06, Podolsky_prb09}.

\begin{acknowledgments}
We thank Joon Hyun Kim and Deokhwa Hong for experimental assistance, and H. Ishihara for useful discussion. This work was supported by the Samsung Science and Technology Foundation (Project No. SSTF-BA1601-06), the Institute for Basic Science in Korea (Grant No. IBS-R009-D1),  and the Japan Society for the Promotion of Science (KAKENHI Grant No. JP17K05549), and in part by the Osaka City University Strategic Research Grant 2017 for young researchers.
\end{acknowledgments}

\clearpage
\newpage

\begin{center}
\textbf{\large Supplemental Material}
\end{center}

\setcounter{equation}{0}
\setcounter{figure}{0}
\setcounter{table}{0}
\makeatletter
\renewcommand{\theequation}{S\arabic{equation}}
\renewcommand{\thefigure}{S\arabic{figure}}
\renewcommand*{\bibnumfmt}[1]{[S#1]}
\renewcommand{\thesubsection}{A\arabic{subsection}}

\def\beq{\begin{eqnarray}} \def\eeq{\end{eqnarray}}
\def\bea{\begin{eqnarray}} \def\eea{\end{eqnarray}}
\def\bse{\begin{subequations}} \def\ese{\end{subequations}}

\def\vecx{{\bm x}} \def\vecy{{\bm y}} \def\vecz{{\bm z}} \def\vecr{{\bm r}} \def\vecv{{\bm v}}\def\vecD{{\bm \nabla}}
\def\hatx{\hat{\bm x}} \def\haty{\hat{\bm y}} \def\hatz{\hat{\bm z}} \def\hatp{\hat{\bm p}} \def\hatk{\hat{\bm k}}
\def\||{\parallel}
\let\p\partial

\def\<{\left\langle} \def\>{\right\rangle}
\def\({\left(} \def\){\right)}
\def\[[{\left[} \def\]]{\right]}

\subsection{Domain wall in the polar phase}

\begin{figure}[b]
	\centering
	\includegraphics[width=8.0cm]{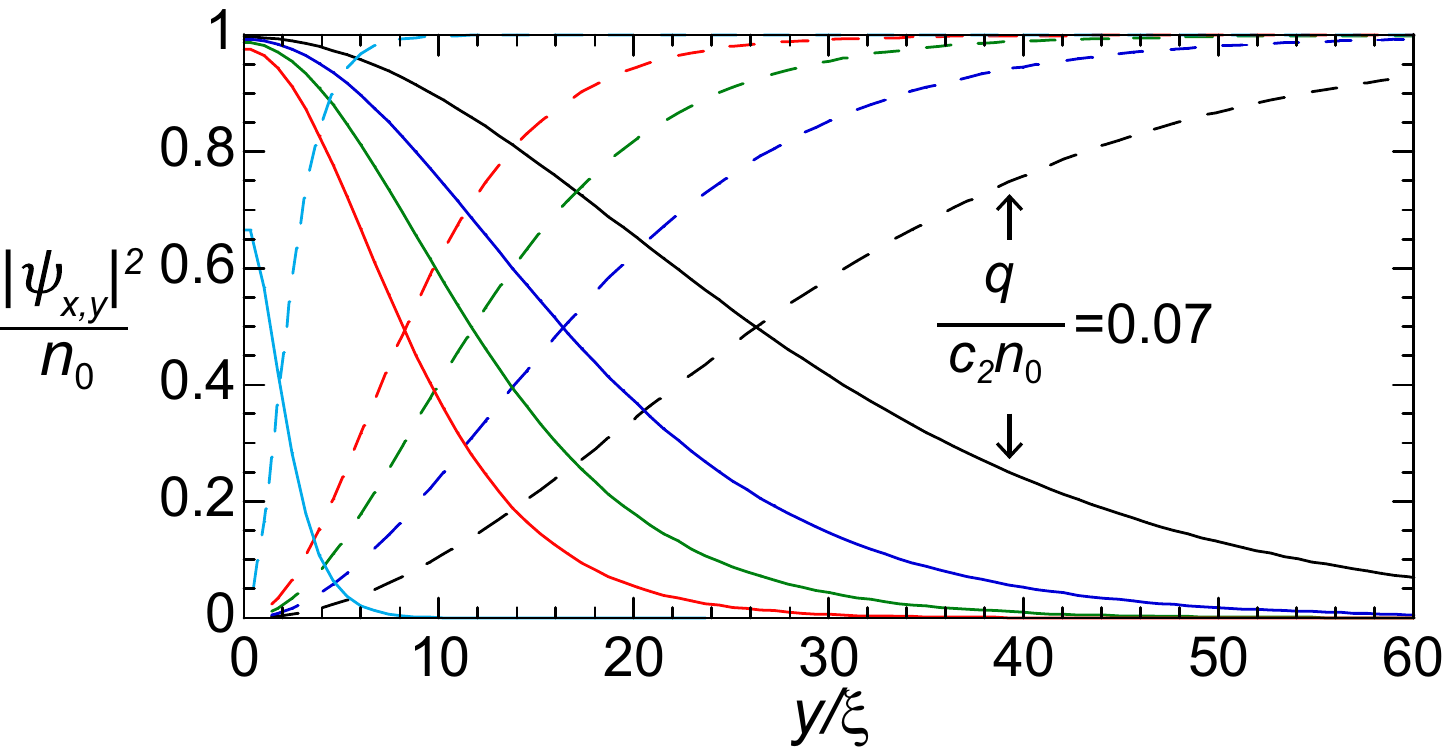}
	\caption{Numerical solutions for planar domain walls. Solid and dashed curves show the spatial profiles of $|\psi_x|^2$ and $|\psi_z|^2$, respectively, for $\frac{q}{c_2n_0}=$ 0.07, 0.18,0.35, 0.7, and 10 (from right to left) with $\frac{c_2n_0}{\mu}=0.016$. Here, $n_0=\mu/c_0$ is the bulk density. The profile is symmetric with respect to $y=0$.
		The wall thickness is characterized by $\xi_q=\frac{\hbar}{\sqrt{2mq}}=\xi\sqrt{\frac{\mu}{q}}$ for $\xi_q \ll \xi$.
		In the core regime ($|y| \lesssim \xi_q $), the total density $n=\left| \psi_x\right |^2+\left| \psi_z\right |^2$ with $\psi_y=0$ is almost constant for $\xi_q \ll \xi$, while it decreases substantially for $\xi_q\sim \xi$ and finally vanishes when $\xi_q/\xi$ reaches to a certain value.}
	\label{fig:NumWall}
\end{figure}

To describe the theme of spontaneous symmetry breaking in our system, we introduce a complex vector field in the Cartesian representation for spin-1 spinor BECs as
\begin{eqnarray}
	{\bm\psi}&=&\left[\psi_x,\psi_y,\psi_z \right]^{\rm T}	\nonumber \\	
	&=&\left[-\frac{1}{\sqrt{2}}\left( \psi_{+1}-\psi_{-1} \right), -\frac{i}{\sqrt{2}}\left( \psi_{+1}+\psi_{-1} \right),\psi_0\right]^{\rm T}.	\nonumber
\end{eqnarray}
In this representation, the energy density ${\cal E}({\bm \psi})$ is written as
\begin{eqnarray}
	{\cal E}({\bm \psi})&=&\frac{\hbar^2}{2m}\sum_{\nu=x,y,z}\left|  \partial_\nu {\bm \psi} \right|^2
	+ q\left( \left| \psi_x \right|^2+\left| \psi_y \right|^2\right)		\nonumber \\
	&&-\mu \left| {\bm\psi} \right|^2
	+\frac{c_0}{2}\left| {\bm\psi} \right|^4
	+\frac{c_2}{2}\left|{\bm\psi}\times {\bm\psi}^*\right|^2.
	\label{eq:energy_density}	
\end{eqnarray}
In antiferromagnetic BECs with $c_2>0$, the ground state is the polar state, where we have the spin density vector ${\bm S}\equiv i {\bm \psi}\times {\bm \psi}^*=0$ and the order parameter of the polar phase represented by
\begin{eqnarray}\label{eq:psi_Td}
	{\bm \psi}=\psi_{\rm T}\hat{\bm d}
\end{eqnarray}
with a complex scalar field $\psi_{\rm T}=\sqrt{n}e^{i\phi}$ and a real unit vector field $\hat{\bm d}=[ d_x, d_y, d_z]^{\rm T}$.

\begin{figure}[b]
	\centering
	\includegraphics[width=\linewidth]{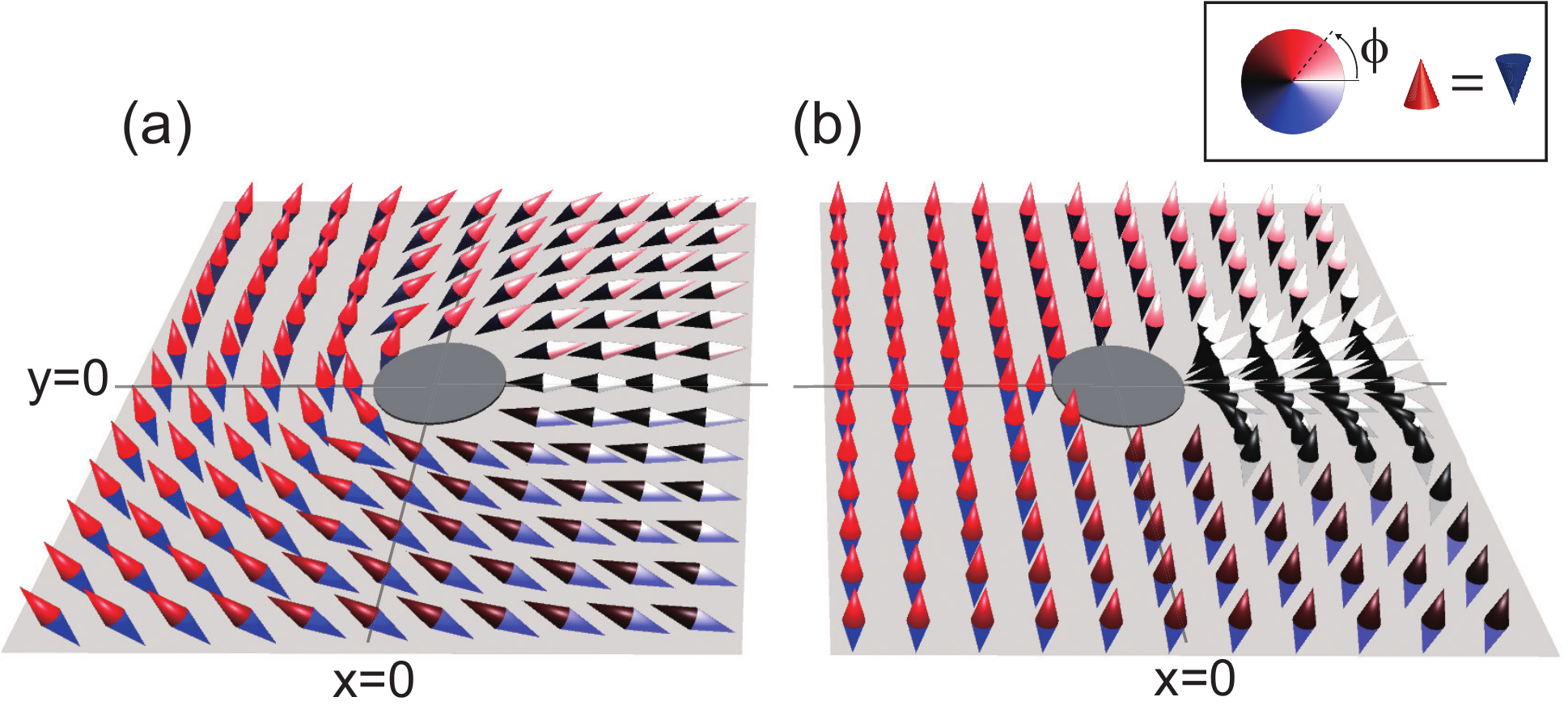}
	\caption{Schematic illustration of (a) a linear HQV and (b) a wall-vortex composite defect. The structure of the composite defect can be described as a deformation from that of the linear HQV, where the spin-rotating region is condensed along the $x$ axis for $x>0$  to form a domain wall terminating at the HQV at the origin.}
\label{fig:HQVs}
\end{figure}

To capture the feature of domain walls, we consider a stationary solution for a planar domain wall perpendicular to the $y$ axis in a uniform system; $\hat{\bm d}\to \pm \hat{\bm z}$ for $y\to \pm \infty$ [Fig.~1(a)]. The vector $\hat{\bm d}$ varies continuously across the wall, with $\hat{\bm d}\bot \hat{\bm z}$ at the core ($y=0$).
We assume $\phi=0$ and $\hat{\bm d}\bot \hat{\bm y}$ for $\psi_{+1}=-\psi_{-1}$ without a loss of generality.
Then, the planar wall is evaluated by the reduced energy density for the real functions $\psi_x$ and $\psi_z$,
\begin{eqnarray}
	{\cal E}&\to&\frac{\hbar^2}{2m}\left[ \left(\frac{d \psi_x}{dy}\right)^2+ \left(\frac{d \psi_z}{dy}\right)^2 \right] 	\nonumber \\
	&&+ q\psi_x^2
	-\mu  \left( \psi_x^2+\psi_z^2 \right)
	+\frac{c_0}{2}\left( \psi_x^2+\psi_z^2 \right)^2.
	\label{eq:energy_density_soliton}
\end{eqnarray}
For the case of $\mu \gg q$, the total density $n$ ($=\psi_x^2+\psi_z^2$) is approximately constant; the wall is thus characterized by the parameter $q$.
Comparing the gradient term and the quadratic Zeeman term,
the thickness of a domain wall is characterized by
\begin{eqnarray}
	\xi_q=\frac{\hbar}{\sqrt{2mq}}.
	\nonumber
\end{eqnarray}
Figure \ref{fig:NumWall} shows numerical solutions for the domain wall for different values of $q$. According to the expression for $\xi_q$, the domain wall becomes thicker as $q$ decreases. This behavior is qualitatively consistent with our experimental observations.

\begin{figure*}[t]
	\centering
	\includegraphics[width=14.5cm]{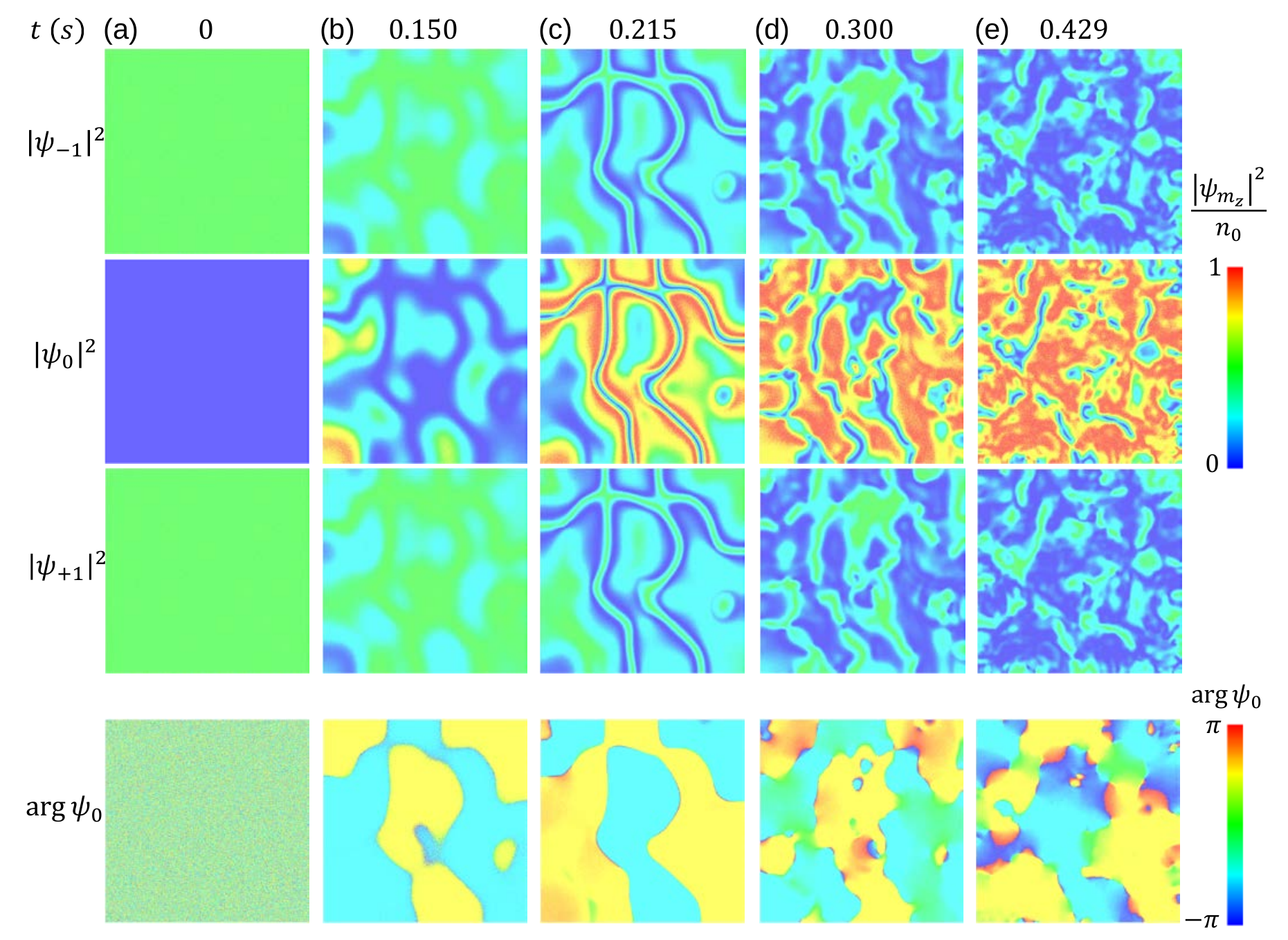}
	\caption{Numerical time evolution of the density $\left|\psi_{m_z=0, \pm 1}\right|^2$ and the phase $\arg{\psi_0}$ for $q_f=2.6$ Hz at $t=$ (a) 0, (b) 0.150, (c) 0.215, (d) 0.300, (e) 0.429 sec. $\mu/h=927$~Hz and $c_2/c_0=0.016$. At $t=0$, $(\psi_0,\psi_{\pm 1})=(0,\pm \sqrt{\frac{n_0}{2}})$.  The size of each box is $724\xi \times 724\xi$. }
	\label{fig:QuenchDynamics}
\end{figure*}

\subsection{Structure of the wall-vortex composite defect}

The structure of a wall-vortex composite defect is understood in a similar manner to that reported in the literature for superfluid $^3$He-B~\cite{Kondo_prl92,Eltsov_prl00}.
In the $^3$He-B system, the composite defect is formed due to the existence of two different length scales; 
the coherence length and the dipolar healing length.
The length of the former is defined by the superfluid condensation energy and is much shorter than that of the latter, which is related to the much weaker spin-orbit interaction. Correspondingly, our system of anti-ferromagnetic spinor BECs has also two different length scales; $\xi=\frac{\hbar}{\sqrt{2m\mu}}$ and $\xi_q$.
The length $\xi$ is defined by the term of $c_0$ (or $\mu$) in  Eq.~(\ref{eq:energy_density}), which is associated with the condensation energy.
The large length $\xi_q$ is related to the much weaker quadratic Zeeman energy with $q\ll \mu$; $\xi \ll \xi_q$.
The length (energy) hierarchy supports the coexistence of different kinds of topological defects as composite defects.
This argument is connected with the condition $q \ll \mu$ to stabilize the domain wall as a part of the composite object.

The two length scales $\xi$ and $\xi_q$ are associated with a vortex and a domain wall, respectively.
If the order parameter varies spatially on length scales much shorter than $\xi_q$,
the quadratic Zeeman term is unimportant compared with the gradient term. 
Then, 
the energy density (\ref{eq:energy_density}) can be reduced to the simplest form
\begin{eqnarray}
	{\cal E} \to\sum_{\nu=x,y,z}\frac{\hbar^2}{2m}\left|  \partial_\nu {\bm \psi} \right|^2
	-\mu \left| {\bm \psi} \right|^2
	+\frac{c_0}{2}\left| {\bm \psi} \right|^4,
	\nonumber
\end{eqnarray}
where again we assume ${\bm \psi} \times {\bm \psi}^*=0$ by considering the polar phase. This form of energy density supports the existence of a HQV. A linear HQV  is realized by considering a combination of two transformations, $\phi \to \phi+ \pi$ and $\hat{\bm d} \to -\hat{\bm d}$, e.g.,
\begin{eqnarray}
	{\bm \psi}(\rho,\theta) =\sqrt{n(\rho)} e^{i\frac{\theta}{2}} \hat{\bm d}(\theta),~~\hat{\bm d}(\theta)=\left[ \cos{\frac{\theta}{2}}, 0, \sin{\frac{\theta}{2}}\right]^{\rm T} 
	\label{eq:HQV_solution}
\end{eqnarray}
with the cylindrical coordinate ${\bm r}=(\rho,\theta,z)$ [Fig.~S2(a)].
Here, the mismatch between $(\phi=0, \hat{\bm d}=\hat{\bm z})$ and $(\phi=\pi, \hat{\bm d}=-\hat{\bm z})$ along the $x$ axis for $x>0$ is avoided and the order parameter varies continuous there because of the discrete symmetry under the operation of $(\phi, \hat{\bm{d}})\rightarrow(\phi+\pi, -\hat{\bm{d}})$.
Note that $\hat{\bm d}$ is ill-defined at the origin because of the discrepancy for its orientation. In the HQV core, the broken-axisymmetry (BA) phase is energetically preferred causing a magnetized vortex core with finite spin density of ${\bm S}\bot \hat{\bm{z}}$.
The size of the magnetized core for the BA phase is at most on the order of the spin healing length $\xi_s=\xi\sqrt{\frac{c_0}{c_2 }}$.

To understand the structure of a wall-vortex composite defect,
it is instructive to show a way to build a composite defect by {\it putting} a HQV in the EAP state.
A HQV is a high-energy object and should be deformed due to the quadratic Zeeman energy with $q>0$.
The order parameter field in Eq.~(\ref{eq:HQV_solution}) is deformed to {\it decrease the area} of the region for $d_x \neq 0$ since  the energy density there is higher than that in the $\hat{\bm d}=\pm \hat{\bm z}$ region.
The phase $\phi$ is not affected directly by the quadratic Zeeman energy.
After the deformation, the $\hat{\bm d}\bot \hat{\bm z}$ region {\it condenses} along the $x$ axis for $x>0$ to form a domain wall terminating at the HQV at the origin [Fig.~S2(b)].
Inside the wall core, the vector field $\hat{\bm d}$ flips in a continuous manner from $\hat{\bm d}=\hat{\bm z}$ to $\hat{\bm d}=-\hat{\bm z}$ with the phase $\phi$ fixed approximately.
The phase $\phi$ rotates from $0$ to $\pi$ about the HQV core, while the vector field $\hat{\bm d}$ is fixed to point along the $z$-direction outside the wall core.

\begin{figure}[b]
	\centering
	\includegraphics[width=7.8 cm]{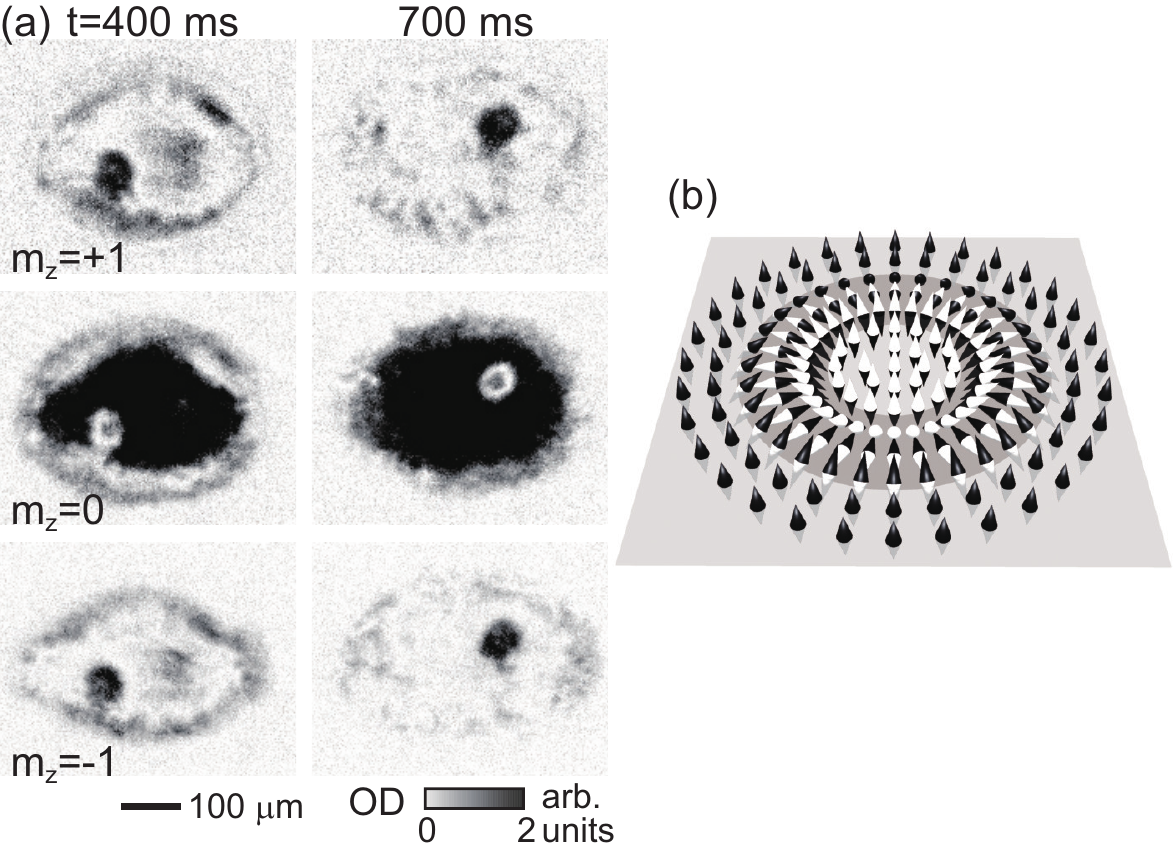}
	\caption{Ring-shaped domain walls in the quenched BECs for $q_f/h$=1.0~Hz. The experimental condition is same as that in Fig.~2. The $|0\rangle$ component shows a density-depleted region of ring geometry, which is filled by the $|$$\pm$$1\rangle$ components. The spatial distributions of the spin components are suggestive of a 2D Skyrmion spin texture which has the topological charge $Q=\frac{1}{4\pi}\int{dx dy~\hat{\bm d}\cdot(\partial_x \hat{\bm d}\times \partial_y\hat{\bm d})}=1$~\cite{Choi_prl12}.}
	\label{fig:skyrmion}
\end{figure}

\subsection{Numerical time evolution of the quench dynamics in a uniform system}

Such composite defects can be nucleated in the course of our non-equilibrium quench dynamics.
Domain walls nucleated via the Kibble-Zurek mechanism for the $\mathbb{Z}_2$ symmetry breaking can transform into composite defects owing to the snake instability, which causes a local Josephson current and nucleates a pair of HQVs by breaking a domain wall into two segments.

To demonstrate this scenario more clearly in a uniform system,
we numerically solve the Gross-Pitaevskii equation for spin-1 BECs. The numerical simulations were done in a method similar to those of the phase transition dynamics described by the Gross-Pitaevskii equation for multi-component systems~\cite{Takeuchi}. Figure \ref{fig:QuenchDynamics} shows a two-dimensional simulation for $q_f/h=2.6$~Hz from the initial EPP state of $(\psi_0,\psi_{\pm 1})=(0,\pm \sqrt{\frac{n_0}{2}})$ under the Neumann boundary condition at $x=\pm \frac{L}{2}$ and $y=\pm \frac{L}{2}$ with the system size $L=724\xi$.
The time evolution is quite consistent with the two-step scenario of the defect formation and explains well the experimental observations.

\subsection{Superflow detection with Bragg scattering}
\begin{figure}[b]
	\centering
	\includegraphics[width=8.5 cm]{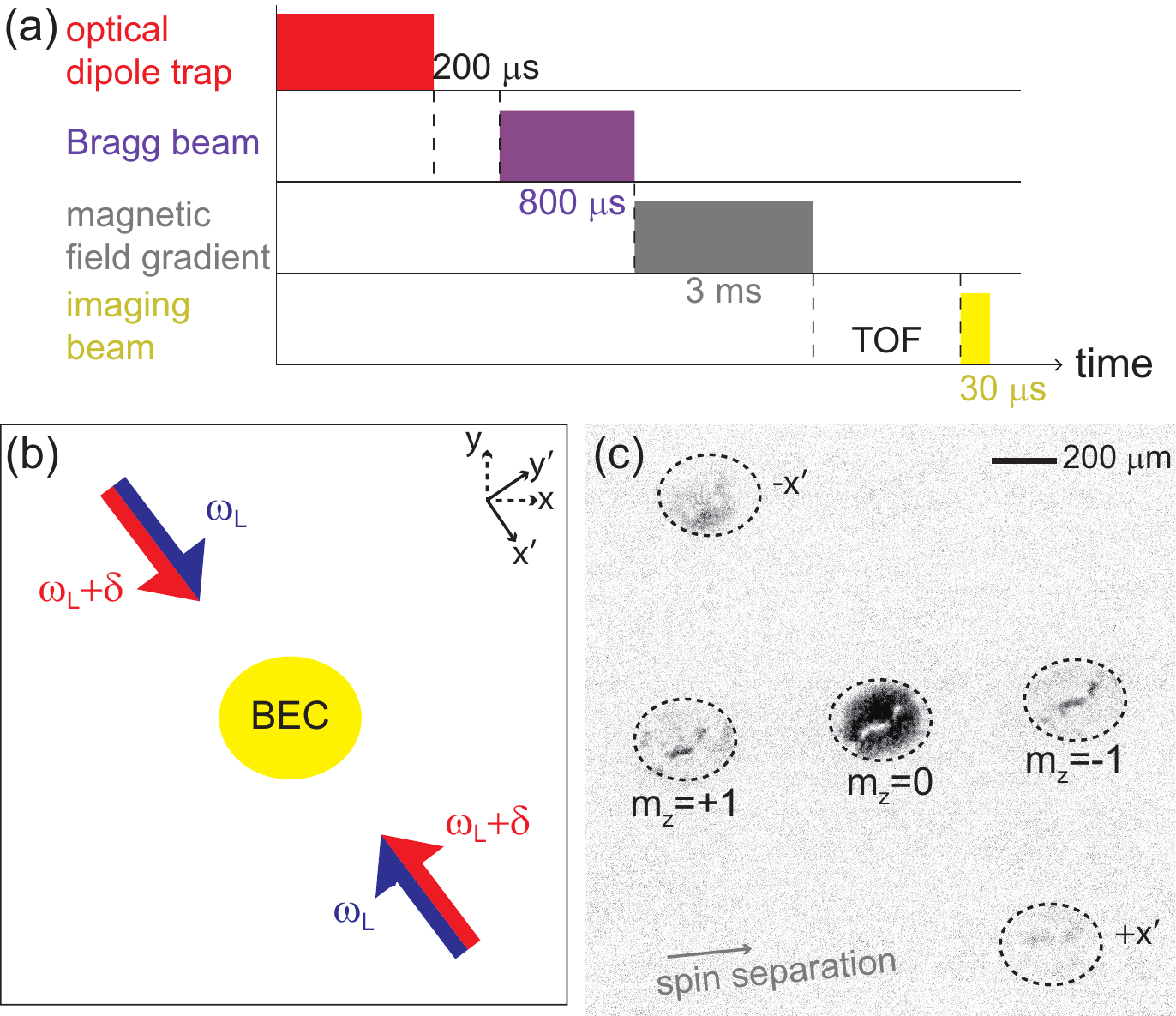}
	\caption{Spatially resolved Bragg spectroscopy with spin separation. (a) Experimental sequence for the Bragg scattering measurement. (b) Schematic of the experimental setup. Two pairs of counterpropagating Bragg beams (blue and red arrows) are irradiated to the sample in the $xy$ plane along the $x'$ axis. After the Bragg beams are applied, a pulse of magnetic field gradient is applied for the Stern-Gerlach spin separation. (c) Example of the image data for $\delta_\text{d}/2\pi=-0.9$~kHz. The three spin components are spatially separated along the field gradient direction (grey arrow) and the two atomic clouds are scattered out from the original $m_z=0$ condnesate along the $\pm x'$-directions, respectively.}
	\label{fig:BraggSGseq}
\end{figure}

The superflow  distribution in the BECs containing composite defects is investigated by employing a spatially-resolved Bragg scattering method~\cite{Seo_scirep17}. When atoms are irradiated by a pair of counterpropagating Bragg laser beams along the $x'$-direction, a two-photon process may resonantly occur by imparting momentum $\vec{p}_0=2\hbar k_L \hat{x}'$ and energy ${\varepsilon}=\hbar \delta$ to the atoms, where $k_L$ is the wavenumber of the two Bragg beams and $\delta$ is their frequency difference. The resonance condition is determined by  $\varepsilon=\frac{p_0^2}{2m}+\vec{p}_0\cdot\vec{v}$ for the atom with velocity $\vec{v}$, giving $\delta_\textrm{d}\equiv \delta-\delta_0=2k_L v_{x'}$ with $v_{x'}=\vec{v}\cdot \hat{x}'$ and $\hbar\delta_0=\frac{p_0^2}{2m}$. Because of the velocity dependence of $\delta$, the spatial distribution of the Bragg scattering response for a BEC directly shows the corresponding velocity region in the BEC.

\begin{figure}[t]
	\centering
	\includegraphics[width=8.5 cm]{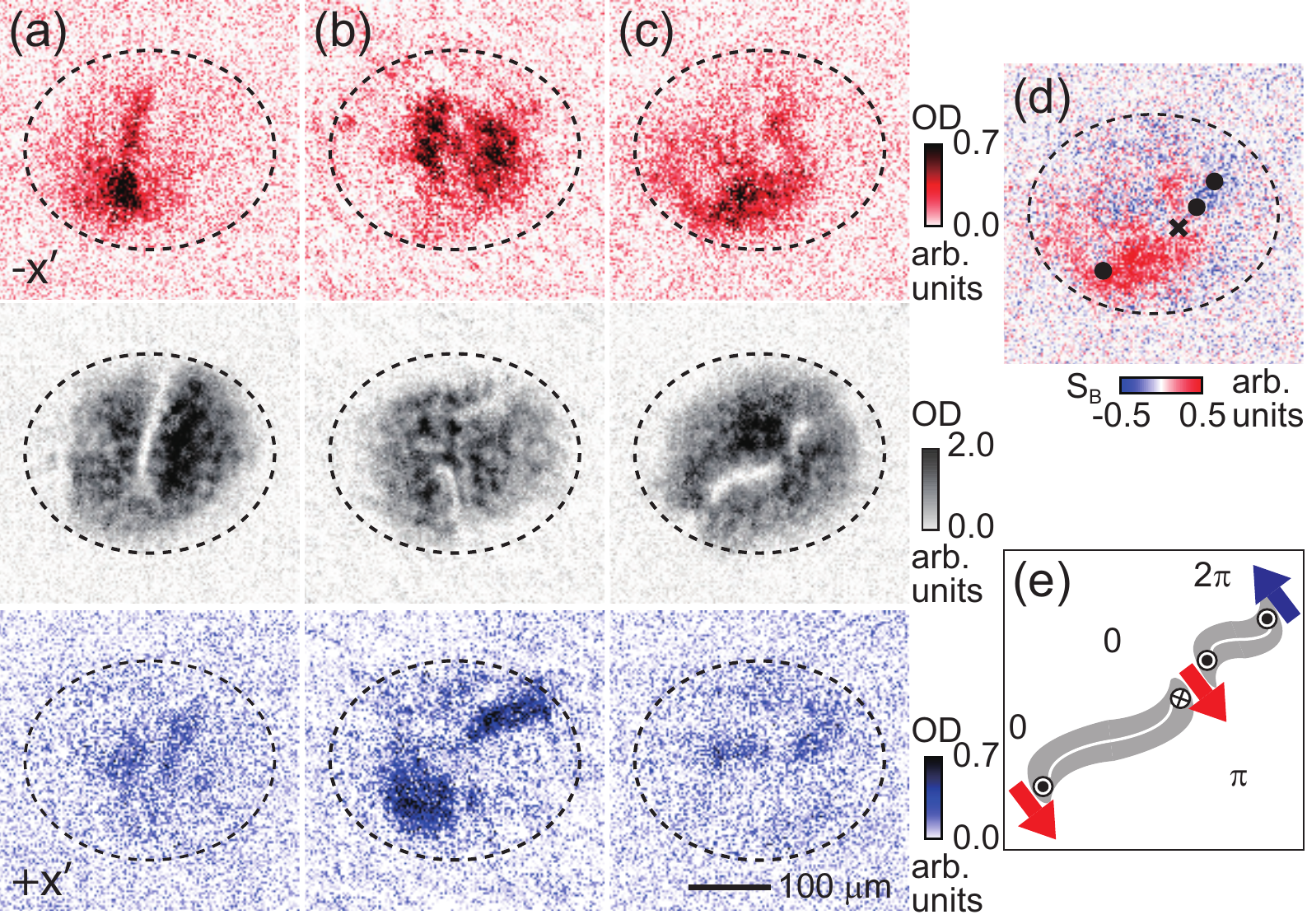}
	\caption{Bragg scattering measurement data for $q_f/h=2.6$~Hz: $-x'$-scattered atom clouds (top), remaining $m_z$$=0$ condensates (middle), and $+x'$-scattered atom clouds (bottom). (a) and (b) are the same data in Fig.~3 and (c) is measured from Fig.~\ref{fig:BraggSGseq}(c). (d) Bragg signal $S_B$ corresponding to (c). The $\bullet$ ($\times$) mark denotes the counterclockwise (clockwise) circulation directions of the HQVs at the wall end points and the dashed lines indicates the boundary of the whole condensate. (e) Schematic description of the defect configuration. The superflow directions are denoted by red and blue arrows.}
	\label{fig:BraggSG}
\end{figure}

Figure~\ref{fig:BraggSGseq}(a) illustrates the experimental sequence for our Bragg scattering measurement. We first turn off the optical dipole trap (ODT). After a short $200~\mu$s time-of-flight (TOF), we apply two pairs of counterpropagating Bragg laser beams to the sample for $800~\mu$s [Fig.~\ref{fig:BraggSGseq}(b)]. Then, we apply an external magnetic field gradient for 3~ms to spatially separate the $|m_z=$$\pm$$1\rangle$ components from the $|0\rangle$ component. After a subsequent TOF, we take an absorption image of the sample including the scattered atom clouds~[Fig.~\ref{fig:BraggSGseq}(c)].  The Bragg signal $S_B(x',y')$ is constructed as $S_B=n_- - n_+$ with $n_\pm (x', y')$ being the density distribution of the atoms scattered out to the $\pm x'$ direction, translated back to the condensate frame~[Figs.~3(c), 3(d), and \ref{fig:BraggSG}(d)]~\cite{Seo_scirep17}. The original density distribution of the $|0\rangle$ component can be obtained by combining that of the unscattered, remaining condensate with $n_+$ and $n_-$, which is used for locating composite defects in the BEC~[Fig.~3(a) and 3(b)].  

\begin{figure}
	\centering
	\includegraphics[width=8.5 cm]{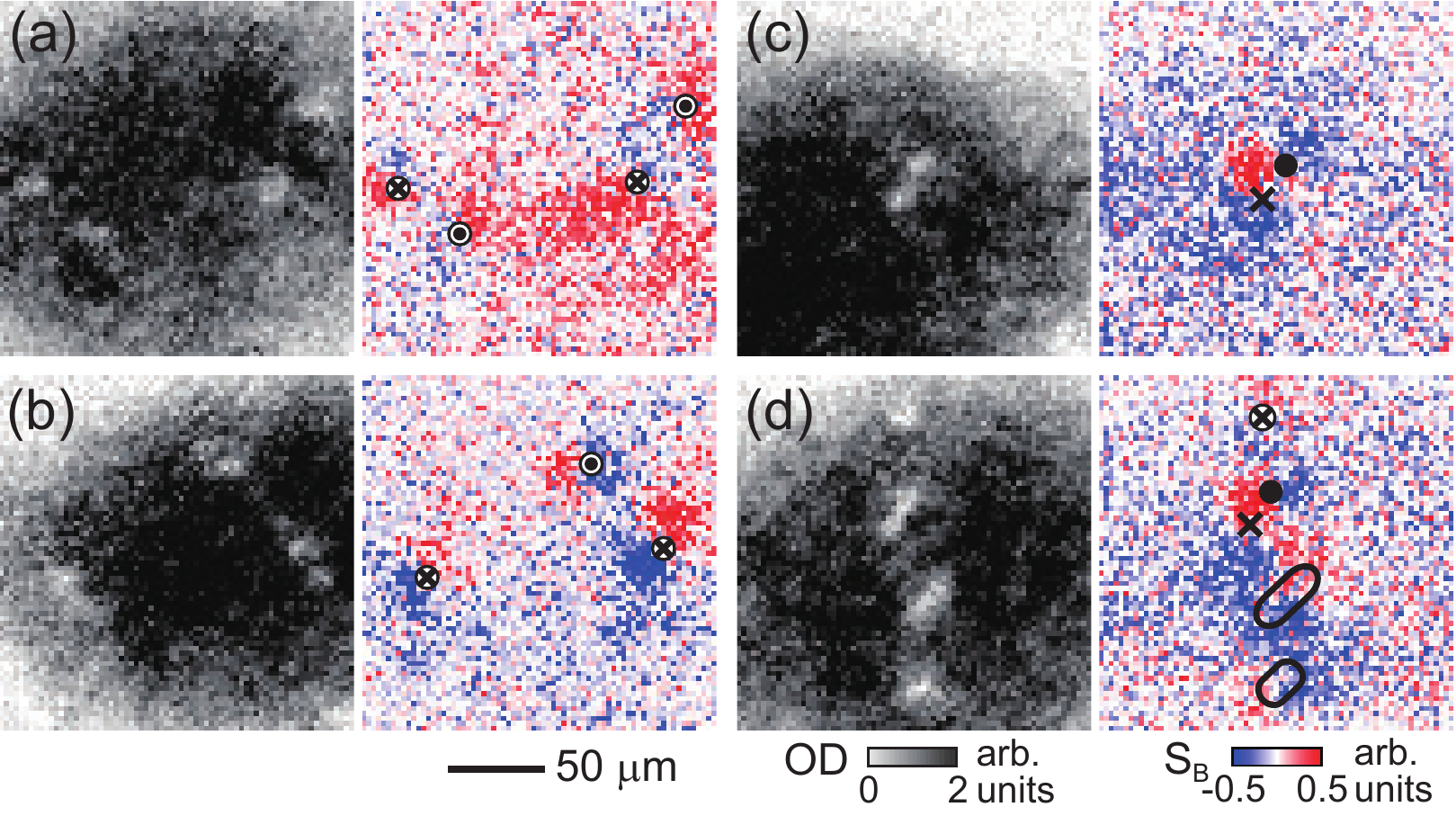}
	\caption{Bragg scattering measurement data for hold times $t=$ (a) 1.2~s and (b-d) 1.5~s with $q_f/h=10.0$~Hz. Absorption images of the remaining $m_z$$=0$ condensates (left) and the corresponding Bragg signals $S_B$ for $\delta_\text{d}/2\pi=-2.0$~kHz (right). In (a,b), $\odot$ ($\otimes$) indicates the circulation direction at the position of round composite defects. (c,d) show the cases of dumbbell-shaped composite defects. Unidentified composite defects are indicated by the open ellipses.}
	\label{fig:RoundDefect}
\end{figure}

In our Bragg scattering experiments, the sample condition is slightly different from that in the experiments described in the main text. The condensate atom number is $N_c\approx 5.8~\times10^6$ and the Thomas-Fermi radii of the condensate are $(R_x, R_y, R_z)\approx(200,143,2.0)~\mu$m for the ODT with trapping frequencies of $(\omega_x, \omega_y,\omega_z)=2\pi\times(4.5,6.3,460)~$Hz. The chemical potential and the peak spin interaction energy are $\mu= h\times 927~$Hz and $c_2 n_0=h\times 14.7~$Hz, respectively. The Bragg beams are red-detuned by $1.7~$GHz from the $|F=1\rangle$ to $|F'=2\rangle$ transition and each beam has an intensity of $0.3~\text{mW/cm}^2$ with a $1/e^2$ width of $1.8~\text{mm}\gg R_{x,y}$. The TOF duration before taking an absorption image was chosen from the range 6 to 11~ms for different measurements.

Figure~\ref{fig:BraggSG}(c) presents one of the rare occasions in the Bragg scattering measurements for low $q_f/h=2.6$~Hz, where the quenched BEC contains free composite defects. Two free composite defects are shown to line up, detached from the boundary of the condensate. The circulation directions of the HQVs at their end points are assigned based on the measured Bragg signal distribution around them~[Fig.~\ref{fig:BraggSG}(d)] despite the signal near the condensate boundary being faint due to low atom density. The free composite defect on the left side shows a strong negative Bragg signal along its wall, indicating its linear motion in the $+x'$-direction, with its moving direction consistent with the circulation directions of its HQVs.

In Fig.~\ref{fig:RoundDefect}, we display four example data of the Bragg scattering measurements at long hold times of $t>1~s$ for high $q_f/h=10.0$~Hz, where several free composite defects of small size are present in the quenched BEC. Taking into account the short distances between the free composite defects, the frequency difference of the Bragg beams was set to be $\delta_\text{d}/2\pi=-2.0~$kHz, corresponding to a higher velocity of $v_{x'}=0.6$~mm/s~$\approx 0.8\frac{\hbar}{m\xi_s}$, to probe a region closer to the HQVs. The Bragg signal around the long-lived, round-shaped defect shows opposite signs at the different lateral sides with respect to the Bragg beam axis line passing through the defect core. This demonstrates the nonzero superflow circulation around the defect, i.e., that the round-shaped defect is of the VV type. We also present a couple of image data for dumbbell-shaped defects in Figs.~\ref{fig:RoundDefect}(c) and (d). The Bragg signal indicates that the defect is of a VAV-type having two HQVs with opposite circulations.


\begin{references}
\bibitem{footnote:Condition} In the Bragg scattering measurements, the sample condition is slightly different. $N_c\approx5.8\times 10^6$, $(R_x, R_y, R_z)\approx(232,160,2.2)~\mu$m, $\mu/h=927~$Hz, and $c_2 n_0/h=14.7$~Hz.
\end{references}
\end{document}